\documentstyle[fleqn,12pt]{article}

\def\pj{\hspace{-.26cm}}
\def\fpj{\hspace{-.7cm}}
\def\thalf{{\textstyle{\frac{1}{2}}}}
\def\frthtw{{\textstyle{\frac{3}{2}}}}
\def\fronsi{{\textstyle{\frac{1}{6}}}}
\def\frfisi{{\textstyle{\frac{5}{6}}}}
\def\frfoth{{\textstyle{\frac{4}{3}}}}
\def\frontw{{\textstyle{\frac{1}{12}}}}
\def\frfitw{{\textstyle{\frac{5}{12}}}}

\def\twothird{{\textstyle{\frac{2}{3}}}}
\def\tquar{{\textstyle{\frac{1}{4}}}}
\def\ttquar{{\textstyle{\frac{3}{4}}}}

\newcommand{\vmg}[1]{\mbox{\boldmath$#1$}}
\begin{document}
\title{An Effective Lagrangian with Broken Scale and Chiral Symmetry III:
\protect\\  Mesons at Finite Temperature}
\author{G.W. Carter, P.J. Ellis and S. Rudaz \\[2mm]
{\small School of Physics and Astronomy}\\[-2mm] {\small University of 
Minnesota, Minneapolis, MN 55455}}
\date{~}
\maketitle
\centerline{{\small\bf Abstract}}
\vskip-.15cm
{\small We investigate the finite temperature behavior of the meson sector 
of an effective Lagrangian which describes nuclear matter. A method is 
developed for evaluating the logarithmic terms in the effective potential which 
involves expansion and resummation; the result is written in terms of the 
exponential integral. In the absence of explicit chiral symmetry breaking,
a phase transition restores the symmetry at a temperature of 190 MeV; when 
the pion has a mass the transition is smooth. At a much higher temperature 
a first order phase transition restores scale symmetry.\hfill\\
PACS: 11.10.Wx, 12.39.Fe\hfill}
\thispagestyle{empty}
\vskip-20cm
\hfill UMN-TH-1523/96; NUC-MINN-96/20-T\\
\newpage
\section{Introduction}

Spontaneous chiral symmetry breaking has long been studied in the linear 
sigma model of Gell-Mann and L\'evy \cite{gml}. At finite temperature the 
restoration of chiral symmetry in both the linear and non-linear versions
of the model has been discussed in, for example, Refs. 
\cite{contre,sasha} (see also references therein). If the potential
of that model, $\tquar\lambda(\sigma^2+\vmg{\pi}^2-f^2)^2$, is used in
calculations of nuclear matter it leads to compression moduli which are much 
larger than the observed value; furthermore the binding energies predicted for
finite nuclei are much too small \cite{glu3,fs2}. In previous work 
\cite{glu3,glu4}, hereinafter referred to as I and II respectively, we have 
therefore replaced this 
potential with a form which incorporates broken scale symmetry in 
addition to spontaneously broken chiral symmetry, as suggested by quantum
chromodynamics (QCD). In particular the potential contains logarithmic 
terms involving the glueball field $\phi$ and the $\sigma$ and $\vmg{\pi}$
fields. At temperature $T=0$ this led to a good description of nuclear 
matter and finite nuclei at the mean 
field level. In II the Lagrangian was extended to include explicit chiral 
symmetry breaking, an additional chiral-invariant term, and the isotriplet 
vector mesons, and satisfactory agreement with low energy $\pi N$ 
scattering data was obtained.

The purpose of the present paper is to examine the finite temperature, 
$T>0$, properties of our Lagrangian. Previous studies \cite{kal,papa} 
of models of this general type at $T>0$ have simply included temperature 
effects for the nucleons. Clearly thermal effects for the mesons
are also needed, particularly those due to the pion which will be dominant 
at low temperatures. However, while the use of 
logarithmic potentials is straightforward 
at $T=0$, it is far from obvious how to proceed at $T>0$, even at the mean 
field level. Since the analysis is quite complicated, we will focus here on 
the mesonic part of our Lagrangian which contains the $\phi$, $\sigma$ 
and $\vmg{\pi}$ fields. The Lagrangian and our thermal analysis is discussed 
in Section 2. We give our numerical results in Section 3 and Section 4 
contains our conclusions.

\section{\bf Theory}
\subsection{\it Equations of Motion}

As mentioned, we simplify the Lagrangian by excluding nucleons, as well 
as the $\omega$ meson which couples to them. We also take the simplest form for
the mesonic contributions from I, augmented by the explicit chiral symmetry 
breaking discussed in II so as to endow the pion with a mass. 
Then our effective Lagrangian 
involves the glueball field $\phi$ and the chiral 
partner fields $\sigma$ and $\vmg{\pi}$ and takes the form
\begin{eqnarray}
{\cal L}_M&\pj=&\pj\thalf\partial_{\mu}\sigma\partial^{\mu}
\sigma+\thalf\partial_{\mu}\vmg{\pi}\cdot\partial^{\mu}\vmg{\pi}
+\thalf\partial_{\mu}\phi\partial^{\mu}\phi-{\cal V}\nonumber\\
{\cal V}&\pj=&\pj B\phi^4\hspace{-.7mm}
\left(\ln\frac{\phi}{\phi_0}-\frac{1}{4}\right)
\hspace{-.73mm}-\hspace{-.73mm}\thalf B\delta\phi^4
\ln\frac{\sigma^2+\vmg{\pi}^2}{\sigma_0^2}
+\hspace{-.74mm}\thalf B\delta \zeta^2\phi^2\!\!\left[\sigma^2+\vmg{\pi}^2
-\frac{\phi^2}{2\zeta^2}\right]\nonumber\\
&&-\tquar\epsilon_1'\left(\frac{\phi}{\phi_0}\right)^{\!2}
\left[\frac{4\sigma}{\sigma_0}-2\left(\frac{\sigma^2
+\vmg{\pi}^2}{\sigma_0^2}\right)-\left(\frac{\phi}{\phi_0}\right)^{\!2}
\,\right]-\ttquar\epsilon_1'\;. \label{lm}
\end{eqnarray}
Here $\zeta=\frac{\phi_0}{\sigma_0}$ and in the vacuum $\phi=\phi_0$, 
$\sigma=\sigma_0$ and $\vmg{\pi}=0$, regardless of whether or not the explicit
symmetry breaking term $\epsilon_1'$ is present (an additional term,
$\epsilon_2'$, was unfavored in II and is omitted here). Thus we have 
spontaneous, as well as explicit, chiral symmetry breaking.
The quantities $B$ and $\delta$ are parameters.
For the latter, guided by the QCD beta function, we take $\delta=4/33$ 
as in I and II. The logarithmic terms here
contribute to the trace anomaly: in addition to the standard contribution
from the glueball field \cite{mig,gomm} there is also a contribution from 
the $\sigma$ field. Specifically the trace
of the ``improved" energy-momentum tensor is
\begin{equation}
\theta^{\mu}_{\mu}=4{\cal V}(\Phi_i)-\sum_i\Phi_i\frac{\delta{\cal V}}
{\delta\Phi_i}=4\epsilon_{\rm vac}\left(\frac{\phi}{\phi_0}\right)^{\!4}\;,
\end{equation}
where $\Phi_i$ runs over the scalar fields \{$\phi,\sigma,\vmg{\pi}$\} and
the vacuum energy, 
$\epsilon_{\rm vac}=-\tquar B\phi^4_0(1-\delta)-\epsilon_1'$.

We take the vacuum glueball mass to be approximately 1.6 GeV
in view of QCD sum rule estimates \cite{shif} of 1.5 GeV and recent lattice 
estimates \cite{svw} of 1.7 GeV.
Since the mass is large in comparison to the temperatures of interest
we shall neglect thermal effects for the glueball.
We define the ratio of the mean field to the vacuum 
value to be $\chi=\phi/\phi_0$. Then Lagrange's equations for the glueball and 
$\sigma$ fields in infinite matter are:
\begin{eqnarray}
&&\fpj0=4B_0\chi^3\ln\chi-B_0\delta\chi\left[2\chi^2\ln\left(
\frac{\sigma^2+\vmg{\pi}^2}{\sigma_0^2}\right) - \left(
\frac{\sigma^2+\vmg{\pi}^2}{\sigma_0^2}\right)+\chi^2\right]
\nonumber\\
&&\qquad-\epsilon_1'\chi\left[\frac{2\sigma}{\sigma_0}-\left(
\frac{\sigma^2+\vmg{\pi}^2}{\sigma_0^2}\right)-\chi^2\right]\;,\nonumber\\
&&\fpj0=-B_0\sigma_0^2\delta\chi^4\left(
\frac{\sigma}{\sigma^2+\vmg{\pi}^2}\right)+B_0\delta\chi^2\sigma
-\epsilon_1'\chi^2(\sigma_0-\sigma)\;, \label{leq1}
\end{eqnarray}
where we have defined $B_0=B\phi_0^4$. 

We wish to take into account thermal effects for the $\sigma$ and 
$\vmg{\pi}$ fields.
To that end we break $\sigma$ into a mean 
field part $\bar{\sigma}$ and a fluctuation $\Delta\sigma$ with mean value
$\langle\Delta\sigma\rangle=0$. The mean value of the pion field 
$\langle\vmg{\pi}\rangle$ is, of course, zero. We write
\begin{eqnarray}
\frac{\sigma^2+\vmg{\pi}^2}{\sigma_0^2}&\pj=&\pj\frac{1}{\sigma_0^2}(
\bar{\sigma}^2+2\bar{\sigma}\Delta\sigma+\Delta\sigma^2+\vmg{\pi}^2)\nonumber\\
&\pj\equiv&\pj \nu^2+2\nu\Delta\nu+\vmg{\psi}^2\;,\label{sigfl}
\end{eqnarray}
where, as in I, $\nu=\bar{\sigma}/\sigma_0$ and here
$\Delta\nu=\Delta\sigma/\sigma_0$ and 
$\vmg{\psi}^2=(\Delta\sigma^2+\vmg{\pi}^2)/\sigma_0^2$.
The treatment of the formal equations (\ref{leq1}) at finite temperature 
is far from obvious. Simply expanding the 
fluctuations out to lowest order, as in Ref. \cite{mannq}, will not 
properly treat $\sigma^2+\vmg{\pi}^2$ when it occurs in the denominator or 
in the logarithm in Eqs. (\ref{leq1}). (It is known that approximating the 
logarithm at zero temperature gives poor results.) We shall proceed in two 
steps. For the first step it is useful to bear in 
mind that at low temperatures $\nu\sim1$ and the thermal average
$\langle\vmg{\psi}^2\rangle$ is small, while at high temperatures $\nu$
is small but the thermal fluctuations are large. This means that the 
cross term $2\nu\Delta\nu$ of Eq. (\ref{sigfl}) is small in both limits.
Thus we expand out this term; odd powers can be 
dropped since the thermal average gives zero. This yields
\begin{eqnarray}
&&\fpj0=B_0\delta\chi\left\langle-2\chi^2\ln(\nu^2+\vmg{\psi}^2)
+\frac{4\chi^2\nu^2\Delta\nu^2}{(\nu^2+\vmg{\psi}^2)^2}
+\frac{8\chi^2\nu^4\Delta\nu^4}{(\nu^2+\vmg{\psi}^2)^4}+\nu^2+\vmg{\psi}^2
\right\rangle\nonumber\\
&&\qquad\qquad\qquad+B_0\chi^3(4\ln\chi-\delta)-\epsilon_1'\chi(2\nu-\nu^2
-\langle\vmg{\psi}^2\rangle-\chi^2)\;,\nonumber\\[2mm]
&&\fpj0=B_0\delta\chi^4\nu\left\langle-\frac{1}{\nu^2+\vmg{\psi}^2}+
\frac{2\Delta\nu^2}{(\nu^2+\vmg{\psi}^2)^2}
-\frac{4\nu^2\Delta\nu^2}{(\nu^2+\vmg{\psi}^2)^3}
+\frac{8\nu^2\Delta\nu^4}{(\nu^2+\vmg{\psi}^2)^4}\right\rangle\nonumber\\
&&\qquad\qquad\qquad+B_0\delta\chi^2\nu-\epsilon_1'
\chi^2(1-\nu)\;. \label{leq2}
\end{eqnarray}
Here the angle brackets indicate that we have taken a thermal average.
For most purposes it is sufficient to truncate at 
${\cal O}(\nu^2+\vmg{\psi}^2)^{-3}$, however in the absence of explicit 
symmetry breaking, $\epsilon_1'=0$, there is a small region near the
chiral phase transition where solutions cannot be obtained. This difficulty 
is alleviated by going one power higher. The expansion parameter is of order 
$4\nu^2\langle\Delta\nu^2\rangle/(\nu^2+\langle\vmg{\psi}^2\rangle)^2$
which {\it a posteriori} we find to be $<0.05$ -- this is satisfactorily 
small. For the thermal average of the square of a fluctuating field,
for example for a component $\pi_a$ of the pion field, one has the
standard result
\begin{equation}
\langle\pi_a^2\rangle=\frac{1}{2\pi^2}\int\limits_0^\infty\,dk
\frac{k^2}{\omega_{\pi}}\frac{1}{e^{\beta\omega_{\pi}}-1}\;. \label{tep}
\end{equation}
Here $\beta=1/T$ is the inverse temperature and 
$\omega_{\pi}^2=k^2+m_{\pi}^{*2}$, with $m_{\pi}^{*2}$ the effective pion 
mass which will be discussed in the next subsection. In calculating 
thermodynamic integrals, such as this, we find it convenient to make use of the 
numerical approximation scheme of Ref. \cite{scott}.

In order to evaluate Eq. (\ref{leq2}) 
we take the second step, which is to make a formal expansion in 
$\vmg{\psi}^2$. It is 
sufficient to consider the logarithm in Eq. (\ref{leq2}) which gives
\begin{eqnarray}
\langle\ln(\nu^2+\vmg{\psi}^2)\rangle&\pj=&\pj\ln\nu^2+
\left\langle\sum_{n=1}^{\infty}\frac{(-1)^{n+1}}{n}\left(
\frac{\vmg{\psi}^2}{\nu^2}\right)^{\!n}\right\rangle\nonumber\\
&\pj=&\pj\ln\nu^2+\sum_{n=1}^{\infty}(-1)^{n+1}(n+1)(n-1)!\,y^{-n}\;.
\label{ln1}
\end{eqnarray}
Here we have defined
\begin{equation}
y^{-1}=\frac{\langle\vmg{\psi}^2\rangle}{2\nu^2}
\equiv\frac{\langle\Delta\sigma^2+\vmg{\pi}^2\rangle}{2\nu^2\sigma_0^2}\;.
\label{ydef}
\end{equation}
In obtaining Eq. (\ref{ln1}) we have used the counting factors described 
in the Appendix. These require that 
$\langle\Delta\sigma^2\rangle =\langle\vmg{\pi}^2\rangle/3$. This will be 
exact in the high temperature regime when chiral symmetry has been 
restored. At lower temperatures where we use the appropriate values for the 
thermal expectation values in Eq. (\ref{ydef}), this will be approximate.
However we do obtain the correct low temperature limit.
We also remark that if the fluctuations in the $\sigma$ field are arbitarily 
ignored, the qualitative behavior is similar to the present case. Thus we 
believe the approximation is reasonable, although we would wish to have 
a more quantitative assessment of the errors involved. It would 
seem to be essential to take into account in some reasonable fashion
the vertices in Eq. (\ref{ln1}) with large numbers of fields attached.

While the series in Eq. (\ref{ln1}) is divergent, we regard it as a formal 
expansion which must be resummed before it can be evaluated. The counting 
factors for an odd number of field components lead to expressions 
involving the error function. For an even number of field components, 
four in the present case, the resummation
requires the exponential integrals \cite{as}
\begin{equation}
E_n(y)=\int\limits_1^{\infty}dt\,\frac{e^{-yt}}{t^n}\;.
\end{equation}
Matching the series of Eq. (\ref{ln1}) to the asymptotic expansion of 
the exponential integrals for large $y$, we obtain
\begin{eqnarray}
\langle\ln(\nu^2+\vmg{\psi}^2)\rangle&\pj=&\pj\ln\nu^2+e^y[E_1(y)+E_2(y)]
\nonumber\\
&\pj=&\pj\ln\nu^2+(1-y)e^yE_1(y)+1\;.\label{ln2}
\end{eqnarray}

In the limit of low temperature ($y\rightarrow\infty$) or high temperature
($y\rightarrow0$) we obtain
\begin{eqnarray}
\langle\ln(\nu^2+\vmg{\psi}^2)\rangle&\pj\rightarrow&\pj\ln\nu^2+
\frac{\langle\vmg{\psi}^2\rangle}{\nu^2}\quad(y\rightarrow\infty)\nonumber\\
&\pj\rightarrow&\pj\ln(0.7631\langle\vmg{\psi}^2\rangle)
\quad(y\rightarrow0)\;.
\end{eqnarray}
The low temperature expression simply corresponds to expanding the logarithm to 
lowest order, as it should. In the high temperature expression the
numerical factor is $0.7631=e^{1-\gamma}/2$, where $\gamma$ is 
Euler's constant.

The remaining quantities needed in Eq. (\ref{leq2}) can be evaluated in 
analogous fashion or, more conveniently, by differentiating Eq. (\ref{ln2})
with respect to $\nu^2$. In order to consistently evaluate the counting we
replace $\Delta\nu^2$ in Eq. (\ref{leq2}) by $\tquar\vmg{\psi}^2$. 
The terms involving a fourth power of the fluctuations are evaluated in 
analogous fashion using the approximation
\begin{equation}
\left\langle\frac{\Delta\nu^4}{(\nu^2+\vmg{\psi}^2)^4}\right\rangle=
3\left\langle\frac{\Delta\nu^2\pi_a^2}{\sigma_0^2(\nu^2+\vmg{\psi}^2)^4}
\right\rangle=
\left\langle\frac{(\vmg{\psi}^2)^2}{8(\nu^2+\vmg{\psi}^2)^4}\right\rangle\;.
\end{equation}
Then Lagrange's equations (\ref{leq2}) can be written in final form:
\begin{eqnarray}
&&\fpj0=B_0\delta\chi^3\biggl[\left(-2+2y-y^2+2y^3+\frthtw y^4+\fronsi y^5
\right)e^yE_1(y)-3+y-\frfisi y^2\nonumber\\
&&-\frfoth y^3 -\frontw y^4-2\ln\nu^2\biggr]
+(B_0\delta+\epsilon_1')\chi\nu^2\left(1+\frac{2}{y}\right)+4B_0\chi^3\ln\chi
\nonumber\\
&&-\epsilon_1'\chi(2\nu-\chi^2)\;,\label{leq3a}\\[2mm]
&&\fpj0=\frac{B_0\delta\chi^4y}{6\nu}\left[(3y^2+6y^3+y^4)e^yE_1(y)-3+y-5y^2
-y^3\right]\nonumber\\
&&+B_0\delta\chi^2\nu-\epsilon_1'\chi^2(1-\nu)\;.\label{leq3b}
\end{eqnarray}
Exact chiral symmetry restoration, $\nu=0$, will only be obtained from
Eq. (\ref{leq3b}) when there is no explicit symmetry breaking, 
$\epsilon_1'=0$. We also note that Eqs. (\ref{leq3a}) and (\ref{leq3b})
permit a solution $\chi=0$ corrsponding to scale restoration, in which case 
$\nu$ is undefined.

\subsection{\it Masses}

For each field we define the effective mass at finite temperature as the 
thermal average 
of the second derivative of the potential. This means that we only consider 
contributions arising from a single interaction vertex. Since the mixing 
between the glueball and the $\sigma$ meson is small, we neglect it here for 
simplicity. Specifically
\begin{eqnarray}
\sigma_0^2m_{\sigma}^{*2}&\pj=&\pj\left\langle\frac{\partial^2{\cal V}}
{\partial\Delta\sigma^2}\right\rangle=(B_0\delta+\epsilon_1')\chi^2  
+\left\langle-\frac{B_0\delta\sigma_0^2\chi^4}{\sigma^2+\vmg{\pi}^2}
+\frac{2B_0\delta\sigma_0^2\chi^4\sigma^2}{(\sigma^2+\vmg{\pi}^2)^2}
\right\rangle\;,\nonumber\\[2mm]
\sigma_0^2m_{\pi}^{*2}&\pj=&\pj\left\langle\frac{\partial^2{\cal V}}
{\partial\pi_a^2}\right\rangle=(B_0\delta+\epsilon_1')\chi^2+\left\langle
-\frac{B_0\delta\sigma_0^2\chi^4}{\sigma^2+\vmg{\pi}^2}
+\frac{2B_0\delta\sigma_0^2\chi^4\pi_a^2}{(\sigma^2+\vmg{\pi}^2)^2}
\right\rangle\;,\nonumber\\[2mm]
\phi_0^2m_{\phi}^{*2}&\pj=&\pj\left\langle\frac{\partial^2{\cal V}}
{\partial\phi^2}\right\rangle=4B_0\chi^2(3\ln\chi+1)+3
(\epsilon_1'-B_0\delta)\chi^2+(B_0\delta+\epsilon_1')\nu^2\nonumber\\
&&\qquad-2\epsilon_1'\nu+\left\langle-6B_0\delta\chi^2
\ln\left(\frac{\sigma^2+\vmg{\pi}^2}{\sigma_0^2}\right)
+(B_0\delta+\epsilon_1')\vmg{\psi}^2\right\rangle\!.\label{mass1}
\end{eqnarray}
As we have remarked, we do not consider thermal fluctuations in the glueball
field and so its mass does not enter the equations. 
However it will be useful to display the mass in Sec. 3. The $\sigma$ and 
$\pi$ masses are needed in evaluating $\langle\vmg{\psi}^2\rangle$ as 
indicated in Eq. (\ref{tep}). As a result the equations of motion and the
expressions for the masses must be evaluated self-consistently.
Treating Eq. (\ref{mass1}) as discussed in the previous subsection and using 
the equations of motion (\ref{leq3a}) and (\ref{leq3b})
for the case where $\nu\neq0$ and $\chi\neq0$ we obtain
\begin{eqnarray}
\sigma_0^2m_{\sigma}^{*2}&\pj=&\pj
2B_0\delta\chi^4\nu^2\left\langle\frac{1}{(\nu^2+\vmg{\psi}^2)^2}
-\frac{8\Delta\nu^2}{(\nu^2+\vmg{\psi}^2)^3}
+\frac{4(3\nu^2\Delta\nu^2+2\Delta\nu^4)}{(\nu^2+\vmg{\psi}^2)^4}
\right\rangle\nonumber\\
&&+\frac{\epsilon_1'\chi^2}{\nu}\nonumber\\
&\pj=&\pj\frac{B_0\delta\chi^4y^2}{3\nu^2}\biggl[-(18y+15y^2+2y^3)e^yE_1(y)
+7+13y+2y^2\biggr]\nonumber\\
&&+\frac{\epsilon_1'\chi^2}{\nu}\;,\label{msig}\\
\sigma_0^2m_{\pi}^{*2}&\pj=&\pj\frac{\epsilon_1'\chi^2}{\nu}\;,\label{mpi}\\
\phi_0^2m_{\phi}^{*2}&\pj=&\pj4(B_0\chi^2+\epsilon_1'\nu)
-2(B_0\delta+\epsilon_1')(\nu^2+\langle\vmg{\psi}^2\rangle)\;.\label{mphi}
\end{eqnarray}
It is straightforward to verify by taking the $y\rightarrow\infty$ limit
that the zero temperature results of II are obtained.
In the case when $\nu,y\rightarrow0$ the 
$\sigma$ and $\pi$ masses become equal:
\begin{equation}
\sigma_0^2m_{\sigma}^{*2}\rightarrow\sigma_0^2m_{\pi}^{*2}=
\frac{\epsilon_1'\chi^2}{\nu}\rightarrow B_0\delta\chi^2
\left(1-\frac{\chi^2}{\langle\vmg{\psi}^2\rangle}\right)\;,\label{mnu0}
\end{equation}
where for the last expression we have used Eq. (\ref{leq3b}).
In the case where there is no explicit symmetry breaking, $\epsilon_1'=0$, 
at sufficiently high temperature $\nu=0$ and the masses are precisely 
given by the last expression in Eq. (\ref{mnu0}). 

\subsection{\it Thermodynamics}

The grand potential per unit volume can easily be written down:
\begin{eqnarray}
\frac{\Omega}{V}&\pj=&\pj\langle{\cal V}\rangle+\frac{T}{2\pi^2}\int dk\,k^2
\left[\ln(1-e^{-\beta\omega_{\sigma}})+3\ln(1-e^{-\beta\omega_{\pi}})\right]
\nonumber\\
&&-\thalf m_{\sigma}^{*2}\langle\Delta\sigma^2\rangle
-\thalf m_{\pi}^{*2}\langle\vmg{\pi}^2\rangle\;,\label{grand}
\end{eqnarray}
where $\omega_{\sigma}^2=k^2+m_{\sigma}^{*2}$ and 
$\omega_{\pi}^2=k^2+m_{\pi}^{*2}$. The subtraction of the last two terms 
in Eq. (\ref{grand}) is 
necessary to avoid double counting \cite{lee}. The evaluation of the 
thermal average of the potential follows the discussion in Subsec. 2.1 and 
we simply quote the result.
\begin{eqnarray}
\langle{\cal V}\rangle&\pj=&\pj\chi^4[B_0\ln\chi-\tquar 
B_0(1+\delta)+\tquar\epsilon_1')] 
+(B_0\delta+\epsilon_1')\chi^2\nu^2\left(\frac{1}{2}+\frac{1}{y}\right)
\nonumber\\
&&-\epsilon_1'\chi^2\nu -\thalf B_0\delta\chi^4\biggl[
(1-y+\thalf y^2-y^3-\ttquar y^4-\frontw y^5)e^yE_1(y)\nonumber\\
&&+1-\thalf y+\frfitw y^2+\twothird y^3+\frontw y^4+\ln\nu^2\biggr]
+\tquar[B_0(1-\delta)+\epsilon_1']\;.
\end{eqnarray}
We have added a constant term here so that $\langle{\cal V}\rangle$
is zero in the vacuum. The pressure $P$ is of course $-\Omega/V$.

Now if one takes the partial derivative of $\Omega/V$ with respect to
$\chi$ or $\nu$ the equations of motion (\ref{leq3a}) and (\ref{leq3b}) 
ought to be obtained. This is true if one ignores the dependence of the
masses on these variables. If this is taken into account derivatives
of the explicit $m^2$ terms in (\ref{grand}) cancel with derivatives of the 
Bose partition functions. Derivatives of $\langle\vmg{\psi}^2\rangle$ 
do not cancel precisely. They would do so if all the terms arising from 
derivatives of the original logarithm  of Eq. (\ref{lm}) were retained in 
the subsequent equations. However this gives additional contributions to 
the pion mass so that in the absence of explicit symmetry breaking the 
mass is no longer zero at low temperatures in violation of Goldstone's 
theorem. As it is necessary to approximate, one cannot have it both ways.
We prefer to truncate the equations of motion and the mass expressions 
at a given order and approximate 
the grand potential-- we have verified that the additional terms needed to
produce the exact equations of motion are small in comparison to the 
terms retained.

In this spirit we ignore the derivatives of $\langle\vmg{\psi}^2\rangle$ 
in deriving the energy density which takes the simple form
\begin{eqnarray}
\frac{E}{V}&\pj=&\pj\langle{\cal V}\rangle
-\thalf m_{\sigma}^{*2}\langle\Delta\sigma^2\rangle
-\thalf m_{\pi}^{*2}\langle\vmg{\pi}^2\rangle\nonumber\\
&&+\frac{1}{2\pi^2}\int dk\,k^2\left[\frac{\omega_{\sigma}}
{e^{\beta\omega_{\sigma}}-1}+\frac{3\omega_{\pi}}
{e^{\beta\omega_{\pi}}-1}\right]\;.
\end{eqnarray}

\section{Results}

The explicit symmetry breaking parameter, $\epsilon_1'$, is chosen to 
yield the vacuum 
pion mass ($\epsilon_1'=\sigma_0^2m_{\pi}^2$) for one set of calculations
and to be zero for another set. In order to fit nuclear matter saturation
in these two cases we find $B_0$ to be (334.9 MeV)$^4$ and (342.6 MeV)$^4$, 
respectively. Based on the results of II we take 
$\sigma_0=110$ MeV and $\phi_0=140.9$ MeV. We stress that in 
evaluating $\langle\vmg{\psi}^2\rangle$ the values of the sigma and pion 
masses from Eqs. (\ref{msig}) and (\ref{mpi}) are used.

In Figs. 1 and 2 we
display the sigma and glueball mean fields as a function of temperature.
Here, and in the subsequent figures, the dashed line refers to 
$\epsilon_1'>0$ and the solid curve gives the $\epsilon_1'=0$ results.
In Fig. 1 the solid line shows a chiral phase transition and the 
dotted curve indicates a region of instability where the pressure
is not maximized. Similarly the dotted and dash-dotted parts of the curves 
in Fig. 2 indicate regions of instability for the scale restoration phase
transition. The pion and sigma masses are displayed in Figs. 3 and 4, 
respectively. 

Consider first the $\epsilon_1'=0$ results (solid curves) where $\nu$ and 
$m_{\sigma}^*$ become zero at the chiral phase transition 
temperature $T_c$. Below $T_c$ the $\nu=0$ solution results in unphysical 
(imaginary) masses. The value of $T_c$ can be obtained by observing where
the masses in Eq. (\ref{mnu0}) become zero, giving 
$T_c=\sqrt{3}\chi\sigma_0$. The effect of the glueball field is rather 
small here since $\chi$ is only slightly less than unity at these temperatures 
(Fig. 2). With our parameters $T_c=187$ MeV.
Fig. 1 indicates a weakly first order phase transition, however we do not 
believe that our approximation scheme is sufficiently accurate to determine 
the order of the transition. As one would expect, the behavior in the 
transition region is sensitive to the prescription used to define the masses.
For example, we have remarked that if the equations are truncated at a lower 
order no solution is obtained for a small region in this vicinity. For 
$T>T_c$ the field $\nu=0$ and the pion and sigma masses are equal. 

There are numerous estimates of $T_c$ in the 
literature. The standard chiral model \cite{sasha} yields 
$T_c=\sqrt{2}f_{\pi}$ which with the pion decay constant $f_{\pi}=93$ MeV
gives $T_c=132$ MeV. Apart from the numerical factor, our value is larger
because $\sigma_0$ has to be greater than $f_{\pi}$ in order to fit nuclei; 
see the discussion in I and II. We remark that, to a good approximation, the 
effect of modifying $\sigma_0$ is simply to scale the temperatures on the 
abcissae of the figures. As with the standard model our $\sigma$ mass is 
zero at the critical temperature. Gerber and Leutwyler \cite{gerb} have studied 
two-flavor quantum 
chromodynamics using an effective chiral Lagrangian to three loop order and 
estimate a value of 190 MeV for $T_c$ with massless quarks.
Finally we mention that the two-flavor lattice QCD 
results of Brown {\it et al.} \cite{brown} show a second order phase 
transition for massless quarks, but this is washed out by any finite mass.
The latter seems to be in line with our dashed curves with $\epsilon_1'>0$
where there is no phase transition and $\nu$ smoothly decreases to a small
value. Symmetry restoration takes place at somewhat higher temperatures 
when the pion mass is finite, as would be expected. Figures 3 and 4 show
that the pion mass smoothly increases from the vacuum value and the 
sigma mass initially drops, but then starts to increase at $T\sim350$ MeV 
to become degenerate with the pion mass.

There is a further interesting feature which occurs at high temperature, 
although we caution that the model may well not be reliable in this region.
Also, physically, deconfinement may have taken place since lattice calculations 
suggest that chiral restoration 
and deconfinement occur at similar temperatures \cite{satz}. Indeed with 
a crude lowest order treatment of gluons and massless quarks, taking the 
bag constant to be $|\epsilon_{\rm vac}|$, we find that the deconfined 
phase is preferred for $T>170$ MeV. In spite of these caveats
we show (Fig. 2) that at a rather high temperature 
the solution with $\chi\sim1$ becomes unstable and $\chi$ drops to zero.
We interpret this as a first order phase transition which restores scale 
symmetry. With the present approximation the transition temperature is
$\sim550$ MeV.
The physics behind this is indicated in Fig. 5 by a qualitative 
plot of $\Omega/V$ as a function of $\chi$ with $\epsilon_1'>0$
(note that each curve 
is normalized to zero at $\chi=0$). At low temperatures,
while there is a $\chi=0$ solution, the minimum corresponds to $\chi\sim1$.
At $T=530$ MeV the depth of these two minima become equal and beyond this
the stable solution is $\chi=0$. The unstable solution is indicated in Fig. 
2 by the dotted ($\epsilon_1'=0$) or dash-dotted ($\epsilon_1'>0$) curve 
out to the point where the minimum and 
maximum of the potential, marked by dots in Fig. 5, coalesce. The $\chi=0$
regime should not be taken too seriously since 
all that remains of $\Omega$ are the thermal partition functions for 
massless $\sigma$ and $\pi$ mesons. Nevertheless we believe that a first 
order phase transition of this type is a general feature of the model.
The critical temperature is rather high here because of the coupling of the 
glueball to the $\sigma$ and $\pi$ mesons; Agasyan \cite{aga} has discussed 
the pure glue sector where the estimated critical temperature is lower.

In Fig. 6 we plot the glueball mass as a function of temperature. (Note 
that when $\chi=0$ we have set $\nu=0$ since it is undefined and is small 
prior to scale restoration.) The glueball mass
begins to drop significantly with decreasing $\chi$ for $T>400$ MeV,
but remains much 
larger than $m_{\sigma}^*$ and $m_{\pi}^*$. Thermal fluctuations are not 
expected to be important, except, possibly, for
the highest temperatures discussed here.

Finally in Fig. 7 we show the pressure versus the energy density with the 
points labelled by the corresponding temperatures. At low temperatures our
pressure is slightly negative since our approximation to the grand
potential results in inaccuracies at the 1\% 
level in the delicate cancellation of large terms. The chiral phase 
transition is just visible for the solid curve in Fig. 7, and beyond this 
temperature interactions become less important and the massless gas 
result $E/V=3P$ is rapidly approached.

\section{Conclusions}

We have discussed the finite temperature behavior of the meson sector of 
an effective Lagrangian with 
which we have successfully described nuclear matter and finite nuclei.
This is non-trivial because of the $\ln(\sigma^2+\vmg{\pi}^2)$ term in 
the effective potential. Our method of handling this term involved 
expansion and resummation of an infinite series with the final result cast 
in terms of the exponential integral.

Our results showed that at sufficiently high temperature the mean value of 
the $\sigma$ field 
became small, signalling chiral restoration. Hitherto it has not been 
appreciated that chiral symmetry is regained in this type of model. In the 
absence of explicit chiral symmetry breaking a phase transition
was obtained at $T_c=187$ MeV; we do not believe that our calculations are
sufficiently accurate to determine the order of this transition. In the 
physical case where explicit chiral symmetry breaking was present and the 
pion had a vacuum mass a smooth restoration of chiral symmetry was found, 
the onset of which occurred 
at a somewhat higher temperature. Well beyond this a first order phase 
transition took place in which the mean glueball field dropped to zero,
implying restoration of scale invariance. This is an
interesting physical feature, but deconfinement is expected to have taken 
place before this temperature is reached and the application of our model
at such a high temperature should be taken with a grain of salt.
One would like to know how these results, particularly chiral restoration,
are affected when nucleon degrees of freedom are present in addition to 
the mesons. This will be the subject of future work.

We thank J. Schaffner-Bielich for useful comments. We acknowledge partial 
support from the Department of Energy under grants No. DE-FG02-87ER40328
and DE-FG02-94ER40823. G.W.C. thanks the University of Minnesota for a 
Doctoral Dissertation Fellowship. A grant for computing time from 
the Minnesota Supercomputer Institute is gratefully acknowledged.

\section*{\bf Appendix. Counting for a General Vertex}

We need to evaluate the thermal average of $(\vmg{\psi}^2)^n$, where 
\begin{equation}
\vmg{\psi}^2=\sum_{i=1}^N\psi_i^2\;,
\end{equation}
which involves the sum over the squares of $N$ fluctuating fields. 
This corresponds to evaluating diagrams of the type shown in Fig. 8 where,
in general, the flower would have $n$ petals.
We assume that the thermal average 
$\langle\psi_i^2\rangle$ is independent of the label $i$, which amounts to 
assuming the masses of the particles involved are the same. 
Because of this assumption the result of taking the thermal average 
of each possible pair of fields at a general vertex can be written
\begin{equation}
\langle(\vmg{\psi}^2)^n\rangle=c_n\langle\vmg{\psi}^2\rangle^n\;,
\end{equation}
where $c_n$ is a number which gives the counting. Consider
$(\vmg{\psi}^2)^{n+1}$. The two additional fields can be averaged together.
Alternatively we can break one of the $n$ original pairs and combine the 
additional fields with them in 2 ways; this requires that the additional 
fields have the appropriate label $i$ leading to a factor of $1/N$.Thus
\begin{eqnarray}
&&\fpj\langle(\vmg{\psi}^2)^{n+1}\rangle=\left(1+\frac{2n}{N}\right)
\langle\vmg{\psi}^2\rangle\langle\vmg{\psi}^2\rangle^n\quad{\rm or}\nonumber\\
&&\fpj c_{n+1}=\left(\frac{N+2n}{N}\right)c_n\;.
\end{eqnarray}
Since $c_1=1$, we obtain
\begin{equation}
c_n=\frac{(N+2n-2)!!}{N!!N^{n-1}}\;.
\end{equation}
For the case at hand, $N=4$, $c_n=(n+1)!/2^n$.

\centerline{{\bf Figure Captions}}

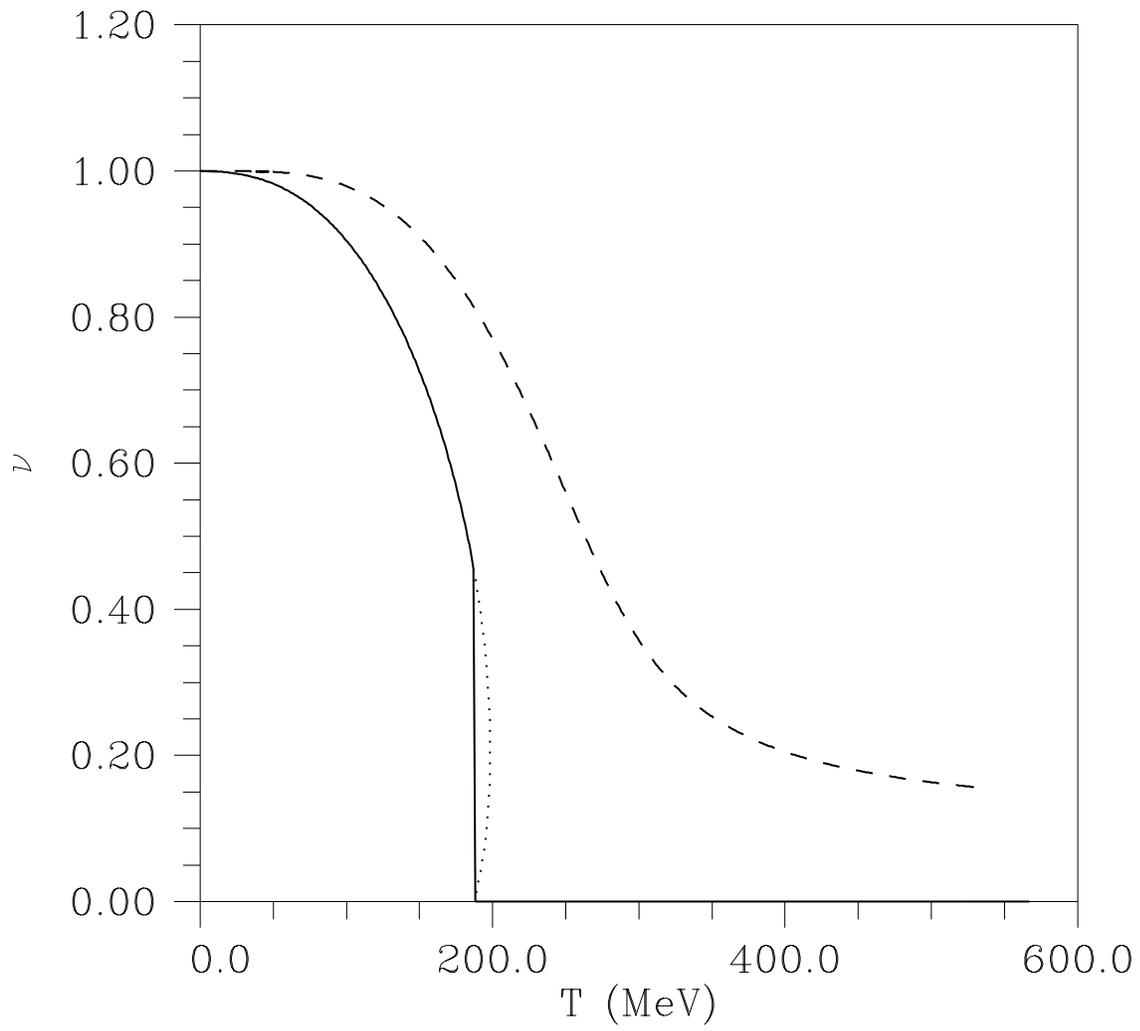
\noindent {\bf Figure 1.} The mean sigma field, $\nu=\bar{\sigma}/\sigma_0$, 
as a function of temperature. The dashed (solid) line corresponds to the 
presence (absence) of explicit chiral symmetry breaking. The dotted curve 
indicates a thermodynamically unstable region.

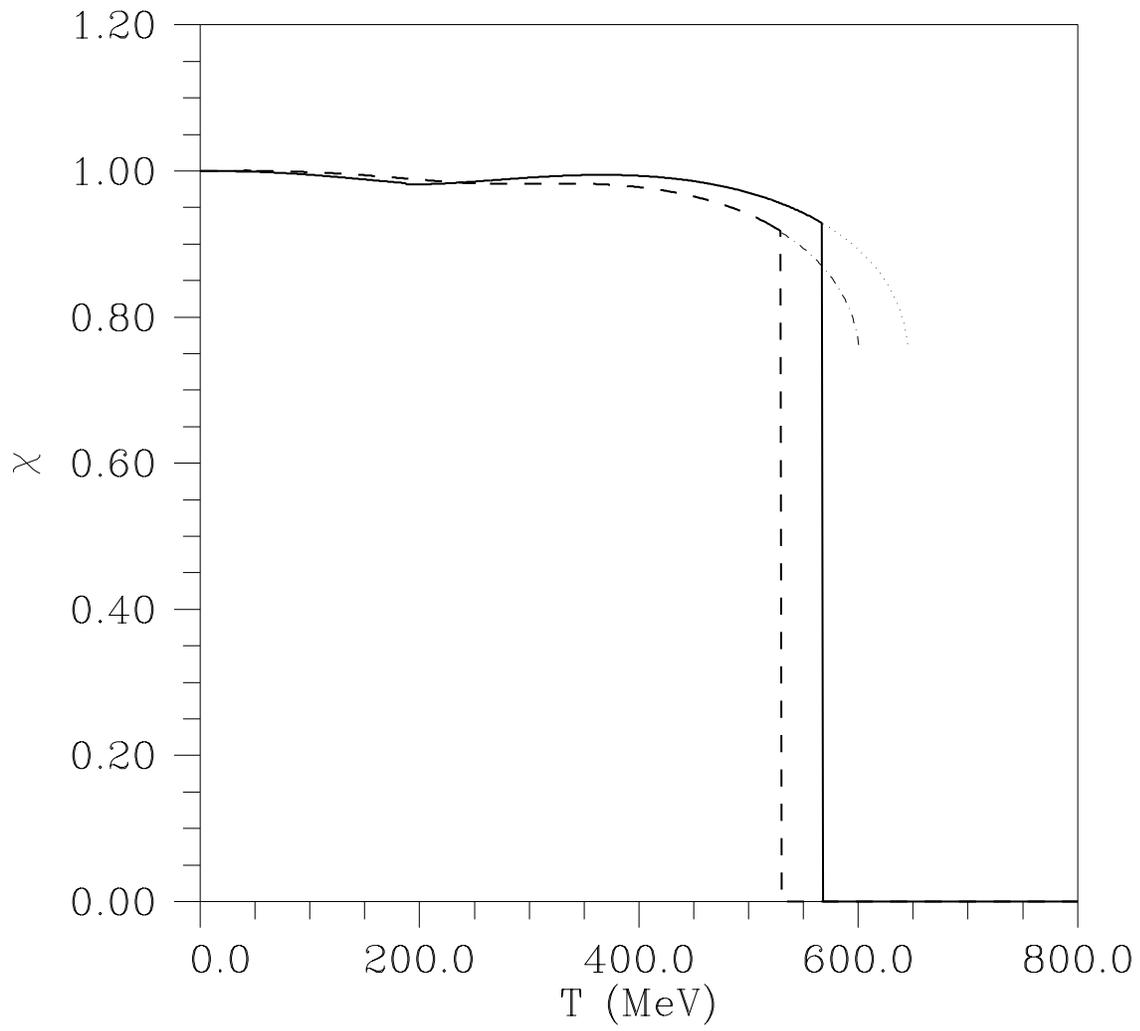
\noindent {\bf Figure 2.} The mean glueball field, $\chi=\phi/\phi_0$, as a 
function of temperature. The dotted and dash-dotted curves indicate 
thermodynamically unstable regions. See caption to Fig. 1.

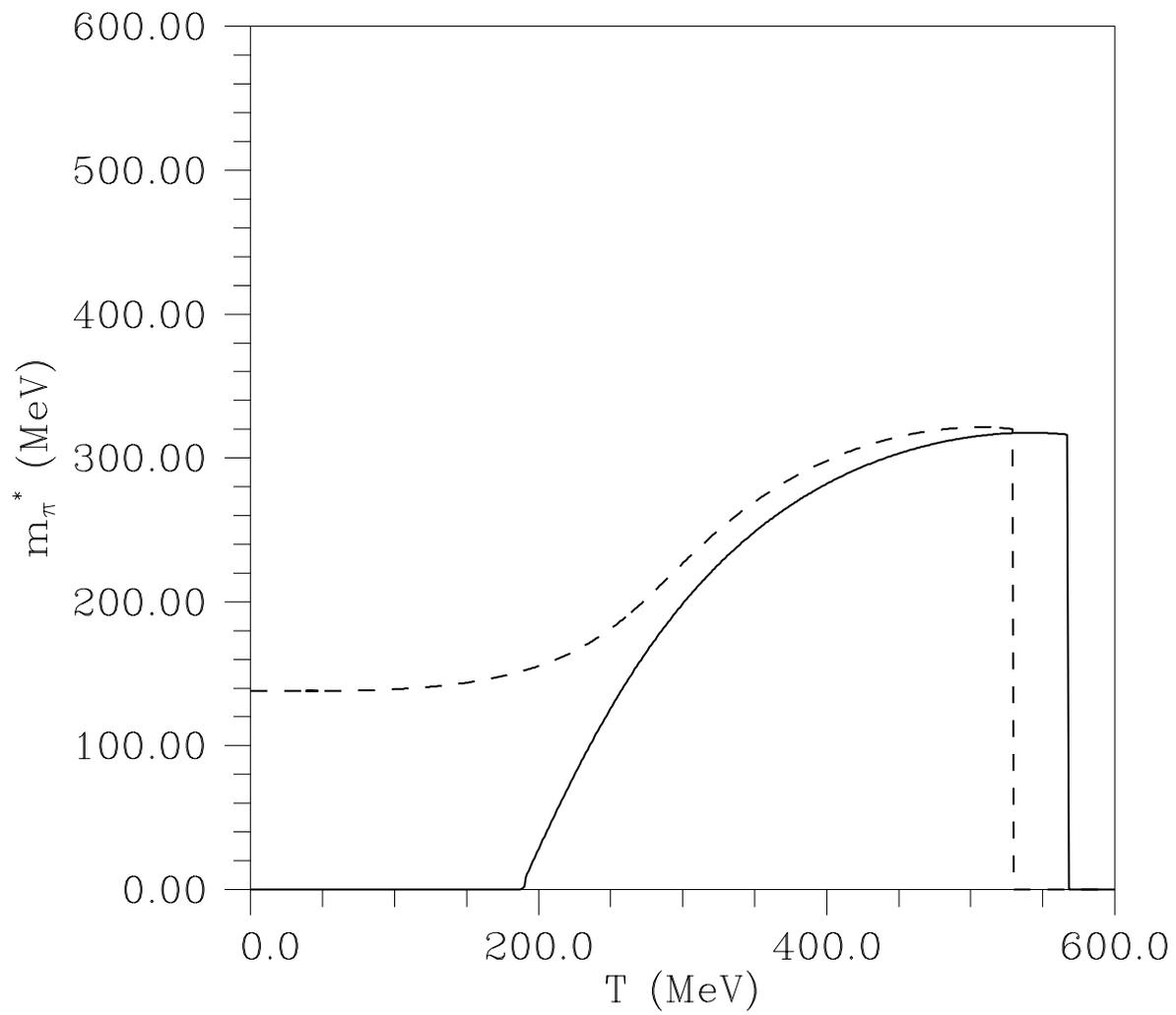
\noindent {\bf Figure 3.} The pion effective mass as a function of temperature.
See caption to Fig. 1.

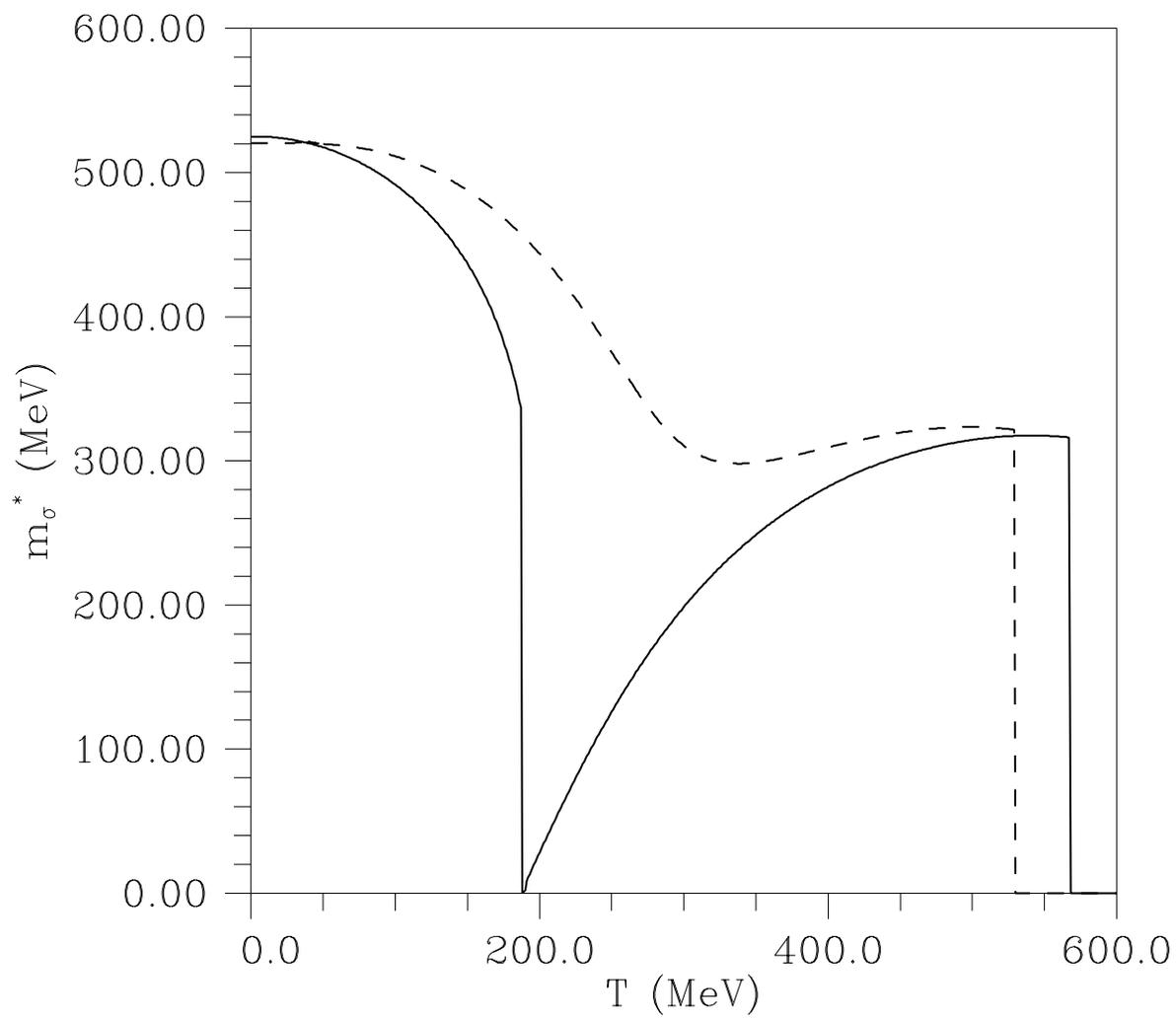
\noindent {\bf Figure 4.} The sigma effective mass as a function of 
temperature. See caption to Fig. 1.

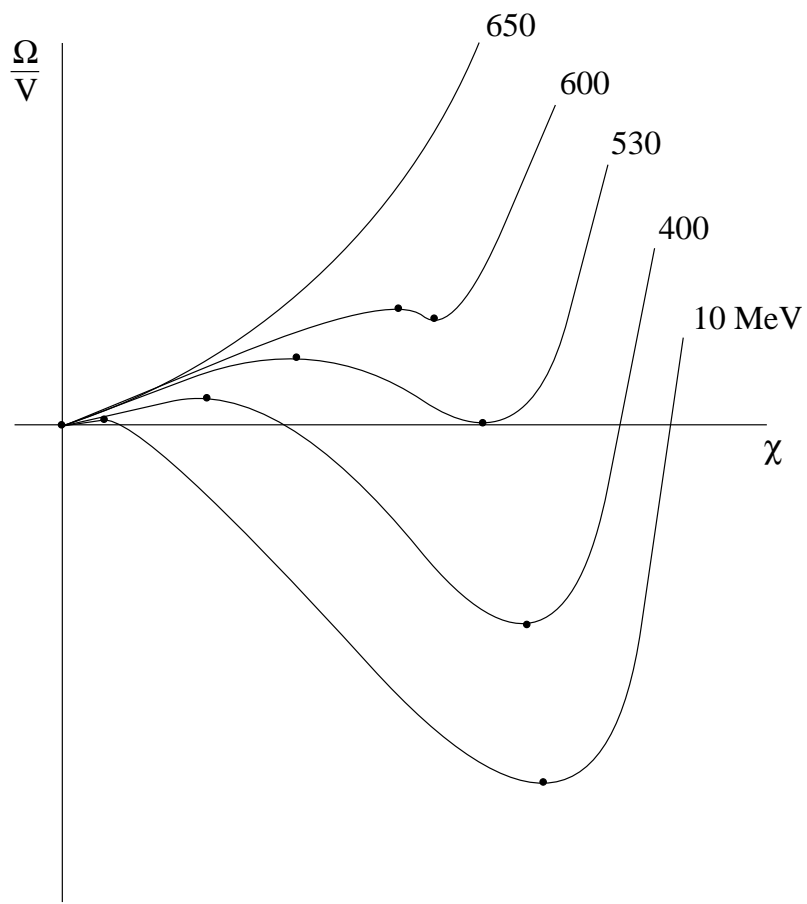
\noindent {\bf Figure 5.} Schematic representation of the grand potential per 
unit volume, $\Omega/V$, as a function of the glueball field $\chi$ for various 
temperatures with $\epsilon_1'>0$. Note that for each case $\Omega$ 
is normalized to zero at $\chi=0$.

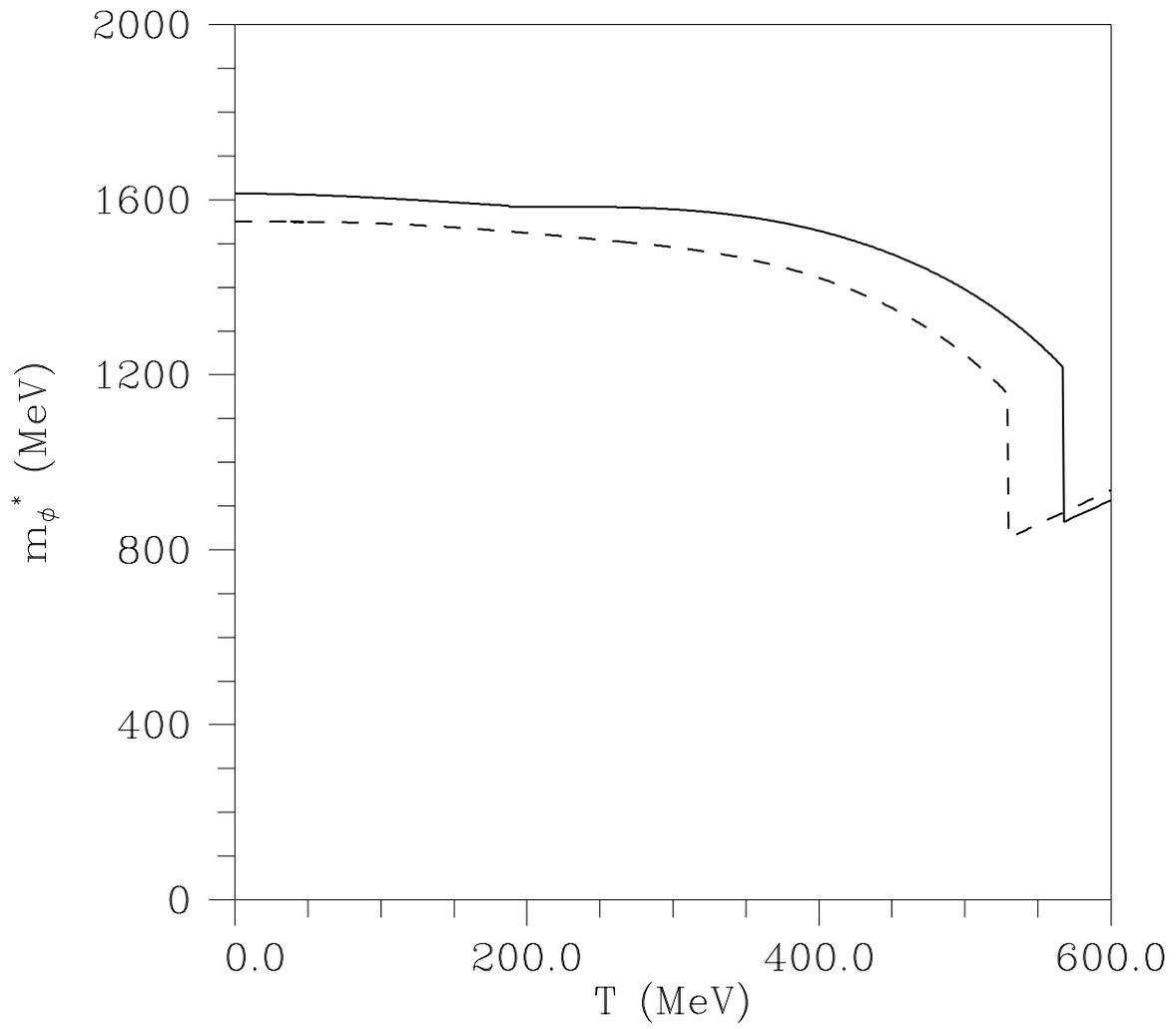
\noindent {\bf Figure 6.} The glueball effective mass as a function of 
temperature. See caption to Fig. 1.

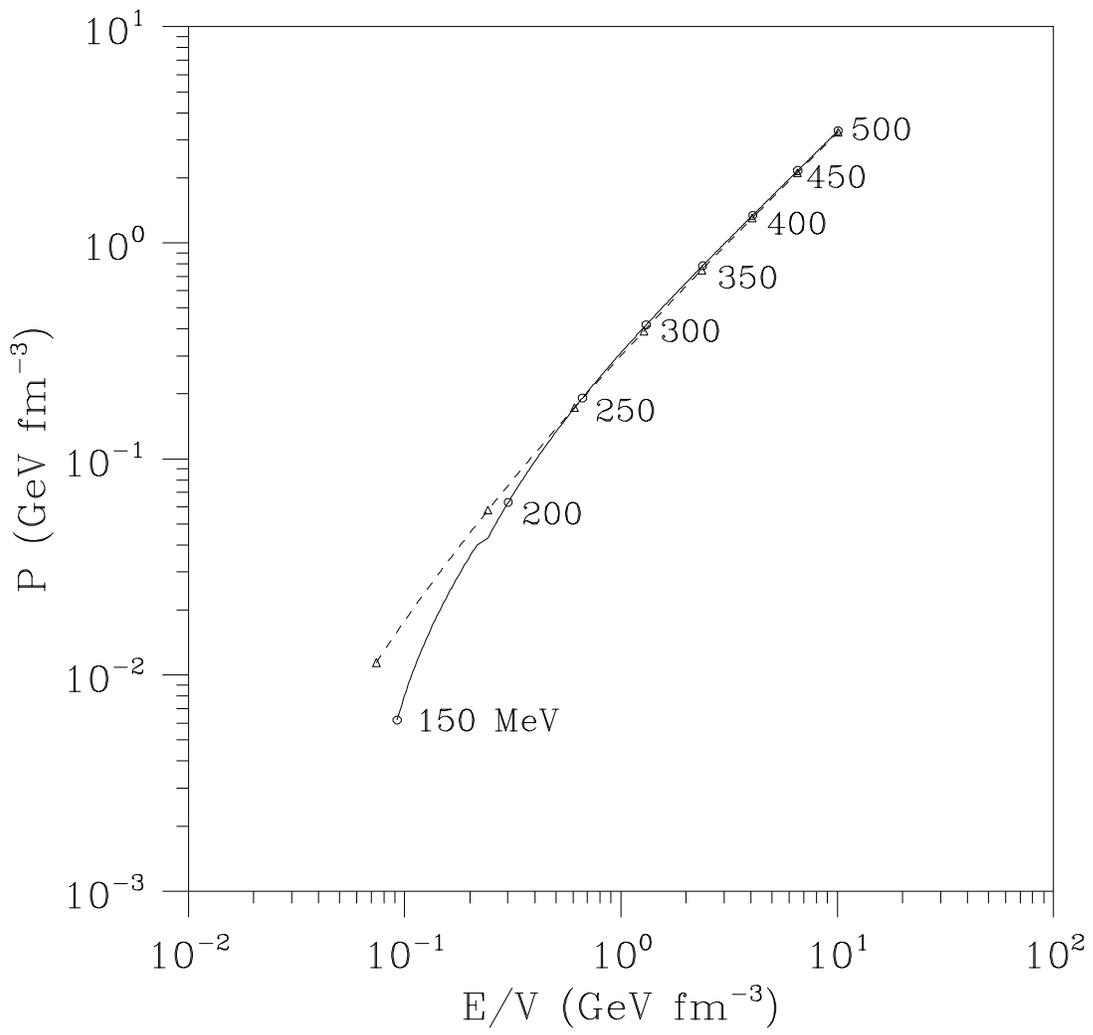
\noindent {\bf Figure 7.} Plot of the pressure versus energy density for the 
temperatures indicated. See caption to Fig. 1.

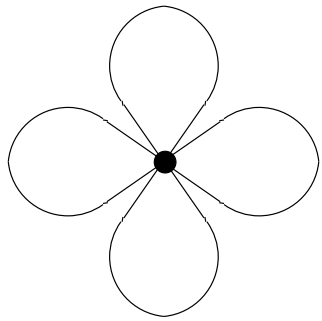
\noindent {\bf Figure 8.} Representative diagram whose thermal part is evaluated 
to determine $\langle(\vmg{\psi}^2)^4\rangle$.
\end{document}